\documentclass[aps,prd,showpacs,preprint]{revtex4}
\usepackage{graphics}
\usepackage{epsfig}
\usepackage{color}
\usepackage{amsmath}

\begin{document}

\newcommand{\be}{\begin{equation}}
\newcommand{\ee}{\end{equation}}
\newcommand{\bea}{\begin{eqnarray}}
\newcommand{\eea}{\end{eqnarray}}
\newcommand{\ba}{\begin{array}}
\newcommand{\ea}{\end{array}}
\newcommand{\nn}{\nonumber}
\newcommand{\slsh}[1]{\not \! #1}

\title{Off-shell Green functions at one-loop level in Maxwell-Chern-Simons quantum electrodynamics}
\author{Y. Concha-S\'anchez$^{1,2}$, A. Raya$^2$ and M. E. Tejeda-Yeomans$^3$} 
\address{$^1$Departamento de Investigaci\'on en F\'{\i}sica, Universidad de Sonora, 
Apartado Postal 5-088, Colonia Centro,
Hermosillo, Sonora 83000, M\'exico. \\
$^2$Instituto de F\'{\i}sica y Matem\'aticas, 
         Universidad Michoacana de San Nicol\'as de Hidalgo \\
         Edificio C-3, Ciudad Universitaria, Morelia, Michoac\'an 58040, M\'exico. \\
$^3$Departamento de F\'{\i}sica, Universidad de Sonora, 
Boulevard Luis Encinas J. y Rosales, 
Colonia Centro, Hermosillo, Sonora 83000, M\'exico.}
\pacs{11.10.Kk,11.15.Bt,11.16.Yc}

\begin{abstract} 
We derive the off-shell photon propagator and fermion-photon vertex at one-loop level in Maxwell-Chern-Simons quantum electrodynamics in arbitrary covariant gauge, using four-component spinors with parity-even and parity-odd mass terms for both fermions and photons. We present our results using a basis of two, three and four point integrals, some of them not known previously in the literature. These integrals are evaluated in arbitrary space-time dimensions so that we reproduce results derived earlier under certain limits.
\end{abstract}
\maketitle

\section{Introduction}

When local gauge symmetry is what drives theories of fermion and boson interactions, mass terms for the latter are forbidden. Thus, the mass of these particles should be explained with different arguments such as the spontaneous breaking of symmetry through a Higgs mechanism.  The recent discovery of a boson with 125 GeV mass and quantum numbers similar to those of the Higgs boson~\cite{LHC} brings this mechanism to a physical reality in the high-energy realm. 
Nonetheless, if leaving local gauge symmetry untouched is desired and yet one wishes to allow the gauge boson to acquire a mass, one could rely on topological quantum field theories such as Chern-Simons (CS) theory~\cite{CS}. The CS theory has been put forward in (2+1)-space-time dimensions in several contexts of condensed matter physics (see, for instance, Ref.~\cite{khare} and references therein), quantum gravity~\cite{witten1} and string theory~\cite{witten2}. The Lagrangian for such a term, although not manifestly gauge invariant, possesses a gauge invariant action. Particularly, considered along with the Maxwell theory for photons, the CS term provides a topological mass for these gauge bosons, which preserves gauge invariance. Moreover by adding matter fields, the theory becomes richer in the sense that some features are apparently unique to (2+1)-dimensional theories, such as anyon excitations, generalized parity, and so on.

In this paper, we study the quantum electrodynamics of four-component fermions and photons in two space and one time dimensions coupled to a CS term, QED$_3$. 
The action of this theory is gauge invariant, though parity and time reversal might be explicitly broken. Fermion and boson mass terms play an intricate role, because in addition to the usual fermion mass term, a parity violating mass term can be included. Such a term would radiatively generate a CS term and vice versa (see, for instance, Refs.~\cite{Edward,Delbourgo}). Specific cases of the underlying Lagrangian have been used in the description of several condensed matter physics systems, once fermions have been integrated out, such as high-$T_{c}$ superconductors~\cite{supercond,supercond2}, the quantum Hall effect \cite{Edward,QHE} and, more recently, graphene~\cite{Gusynin} and topological insulators (see Ref.~\cite{topins} for a recent review). In high-energy physics, the use of QED$_3$ has been connected to the study of dynamical chiral symmetry breaking and confinement, where it provides a popular battleground for lattice and continuum studies \cite{Appelquist}. 
In particular, the CS term allows the possibility of chiral and parity symmetry breaking. Within the two-component fermion formalism, it was observed that while chiral symmetry can be broken, parity symmetry remains untouched~\cite{erik}. Moreover, this term induces a detaching of chiral symmetry breaking-restoration and confinement-deconfinement transitions~\cite{chris1}, leading to a phase resembling quarkonia, where chiral symmetry is restored, but confinement takes place~\cite{chris2}. Confinement and screening effects have  been studied in this theory~\cite{gupta} through the behavior of the vacuum polarization.

A key issue in these type of studies is to address the gauge covariance properties of Green functions. In Ref.~\cite{raya}, the gauge covariance of the fermion propagator was established through the use of the Landau-Khalatnikov-Fradkin transformations~\cite{LKF}. In the same spirit, in Ref.~\cite{Chaichian}, the fermion and photon propagator, and particulary the \emph{on-shell} fermion-photon vertex were calculated and some limits explicitly considered. The same three-point function was studied in Refs.~\cite{Das, lcg}, the later study carried out in the light cone gauge. Interestingly, in all these works, the corresponding anomalous magnetic moment for electrons is identified directly. Here we extend these findings computing the \emph{off-shell} forms of these Green functions in arbitrary covariant gauges. The seminal work of Davydychev {\em et al.}~\cite{davydychev1} on the one-loop quark gluon vertex already provides a guide towards the master integrals involved in the calculation of the two- and three-point Green functions. Moreover, in Refs.~\cite{bashir, Raya-Bashir}, the fermion-photon vertex for the parity preserving version of QED$_3$ was calculated for massless and massive fermions, respectively. Reference~\cite{Raya-Bashir}, in fact, provides the most general form of the non-perturbative fermion-photon vertex in QED3 guided by perturbation theory. In this work, we attempt to generalize these findings incorporating an explicit topological photon mass. In this task, we are faced with new master integrals which were not previously reported in literature, as far as we know. These involve an explicit gauge boson mass, and therefore we evaluate them in arbitrary space-time dimensions, so that these can be useful in other contexts. For us, these steps are fundamental in order to pursue further analysis of certain physical observables sensitive to the influence of the CS term, as will be reported elsewhere.

The paper is organized as follows: In Sec. II, we consider four-component spinors to describe the QED$_{3}$ Lagrangian with parity conserving and violating mass terms for fermions and photons. In Sec. III, we calculate the photon propagator at one-loop level for the most general Lagrangian. The one-loop correction to the fermion propagator is discussed in Sec.~IV. In Sec. V, we derive the fermion-photon vertex at one-loop level under a general decomposition of the vertex we propose. Given this structure of the vertex, new master integrals arise that we calculate and study some of their limiting cases, including the massless case for both fermions and photons. At the end, we summarize our findings in Sec. VI.  Details on the calculation of the new master integrals, are presented in the Appendix.

\section{Maxwell-Chern-Simons QED$_3$}

Let us start by presenting our model and conventions. We work with four-component spinors and thus with a $4\times 4$ representation for the Dirac matrices. We choose to work in Euclidean space, where the Dirac matrices satisfy the Clifford algebra $\{ \gamma_\mu, \gamma_\nu\}=2\delta_{\mu\nu}$. 
Once we have selected a set of matrices to write the Dirac equation, say $\{\gamma_0,\gamma_1,\gamma_2\}$, two anticommuting gamma matrices, namely, $\gamma_3$ and $\gamma_5$, remain unused, leading us to define two independent global phase transformations of the chiral-type for the spinors: $\psi\to e^{i\beta\gamma_3}\psi\;$ and $\psi\to e^{i\sigma\gamma_5}\psi\;,$ where $\beta$ and $\sigma$ are arbitrary real numbers.  Consequently, there exist two types of mass terms for fermions that can be considered in the Lagrangian, the ordinary mass term $m_e \bar\psi\psi$ and a new mass term $m_o \bar\psi\tau\psi\;$ with  
$\tau=\frac{1}{2}[\gamma_3,\gamma_5]$, sometimes referred to as the Haldane mass term~\cite{Haldane} in condensed matter physics. The former violates chirality, whilst the latter is invariant under ``chiral'' transformations. Moreover, $m_e \bar\psi\psi$ is parity invariant but $m_o \bar\psi\tau\psi$ is not. This allows parity and time reversal to be broken in our theory, which is described by
the Lagrangian  
\bea
{\cal L}= \bar\psi (i\slsh{\partial}+e\slsh{A}-m_e-\tau m_o)\psi
-\frac{1}{4}F_{\mu\nu}F_{\mu\nu}-
\frac{1}{2\xi}(\partial_\mu A_\mu)^2 + \frac{\theta}{4}\varepsilon_{\mu\nu\rho}A_{\mu}F_{\nu\rho}, \label{lag}
\eea
where the first term corresponds to the Dirac Lagrangian with parity-even and -odd mass terms and fermion-photon interaction; the next one is the Maxwell term followed by the covariant gauge fixing term, $\xi$ denoting the gauge parameter. The last term is precisely the CS term, with $\theta$ being the topological mass for the photons. Rather than working with parity eigenstates, we find it convenient to work in a chiral basis. For this purpose, we introduce the chiral projectors 
\be
\chi_\pm = \frac{1}{2}(1\pm \tau), \label{chiral projectors}
\ee
which have the properties 
\be
\chi_\pm^2=\chi_\pm\;, \quad
\chi_+\chi_-=0\;,\quad \chi_++\chi_-=1\;. \label{properties projectors}
\ee
Then, the right-handed $\psi_{+}$ and left-handed $\psi_{-}$ fermion fields are $\psi_{\pm}=\chi_{\pm}\psi$, in such a fashion that the chiral decomposition of the fermion propagator becomes \cite{raya}
\be
S_{F}^{(0)}(p)=-\frac{{\not\!p} + m_{+}}{p^2 + m_{+}^2}\chi_{+} - \frac{{\not\! p} + m_{-}}{p^2 + m_{-}^2}\chi_{-}\;,
\ee
with $m_\pm=m_e\pm m_o$. 
Furthermore, we identify the photon propagator associated with the Lagrangian~(\ref{lag}) as
\bea
\Delta_{\mu\nu}^{(0)}(q;\xi)= \frac{1}{q^2+\theta^2}\left(\delta_{\mu\nu}-
\frac{q_\mu q_\nu}{q^2} \right)-\frac{\varepsilon_{\mu\nu\rho}q_\rho \theta}{q^2(q^2+\theta^2)} +\xi \frac{q_\mu q_\nu}{q^4}\;. 
\eea
One can identify from the transverse piece of the propagator the role of $\theta$ as a photon mass. It also plays a role in the parity-odd piece (proportional to $\varepsilon_{\mu\nu\rho}$), but it is absent in the longitudinal part. In the following section we calculate  the one-loop correction to this propagator.

\begin{figure}[t]
\begin{center}
\includegraphics[width=0.25\textwidth]{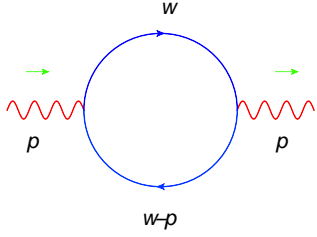}
\end{center}
\caption{One-loop correction to the photon propagator.}
\label{fig1}
\end{figure}

\section{THE ONE-LOOP PHOTON PROPAGATOR}

The one-loop correction to the photon propagator is depicted in Fig.~\ref{fig1}.
Using Feynman rules, it corresponds to the expression
\be
\Pi_{\mu \nu}(p)\,=\,\int\frac{d^3w}{(2\,\pi)^3}
{\it S}^{(0)}_{F}(w)e\gamma_{\mu}
{\it S}^{(0)}_{F}(w-p)e\gamma_{\nu} \;,  \label{propagador}
\ee
which we define in terms in the left- and right-hand side projections as
\be
\Pi_{\mu \nu}(p)=\Pi_{(+)\ \mu\nu}(p) + \Pi_{(-)\ \mu\nu}(p)\;,
\ee
where
\bea
\Pi_{(\pm)\ \mu \nu}(p)=\frac{\alpha}{2\pi^2}\int\frac{d^3w}{(2\,\pi)^3} \frac{[{\not\!w} + m_{\pm}]\gamma_{\mu}[{\not\!w - \not\!p}+ m_{\pm}]\gamma_{\nu}}{[w^2 + m_{\pm}^2][(w -p)^2+ m_{\pm}^2]}\chi_{\pm}\;. \nn 
\eea
To evaluate these expressions, we propose the following decomposition:
\bea
\Pi_{(\pm)\ \mu \nu}(p)&=&
\frac{\alpha}{2\pi^2}\Bigg[K_{\sigma\beta}\gamma_{\sigma}\gamma_{\mu}\gamma_{\beta}\gamma_{\nu} - K_{\sigma}\gamma_{\sigma}\gamma_{\mu}{\not\!p}\gamma_{\nu}
- 2m_{\pm}K_{\sigma}\delta_{\mu\sigma}\gamma_{\nu} -m_{\pm}\gamma_{\mu}{\not\!p}\gamma_{\nu} \nn \\ && \hspace{-3mm} + m_{\pm}^2\gamma_{\mu}\gamma_{\nu}\Bigg]\chi_{\pm}\;,  \label{foton.basis}
\eea
where we have written the propagator in terms of
\bea
K&=&\int d^3w\frac{1}{[w^2 + m_{\pm}^2][(w -p)^2+ m_{\pm}^2]}\;, \nn \\
K_{\mu}&=&\int d^3w\frac{w_{\mu}}{[w^2 + m_{\pm}^2][(w -p)^2+ m_{\pm}^2]}\;, \nn  \\
K_{\mu \nu}&=&\int d^3w\frac{w_{\mu}w_{\nu}}{[w^2 + m_{\pm}^2][(w -p)^2+ m_{\pm}^2]}\;. \label{int. foton}
\eea
These can be evaluated straightforwardly, using standard techniques (see the Appendix for further details and also Ref.~\cite{Yajaira}). Following Ref.~\cite{raya}, we define
\be
I(p,m)=\frac{1}{p}\arctan{\frac{p}{m}}\;,
\ee
so that the basis integrals may be expressed as 
\bea
K&=&\pi^2 I\left(\frac{p}{2},m_\pm \right)\;, \label{K}\\
K_{\mu}&=&\pi^2 \frac{p_{\mu}}{2}I\left(\frac{p}{2},m_\pm \right)\;, \label{Kmu} \\
K_{\mu \nu}&=&\pi^2\delta_{\mu \nu}\bigg[\frac{m_{\pm}}{2} - \bigg(\frac{p^2 + 4m_{\pm}^2}{8}\bigg)
I\left(\frac{p}{2},m_\pm \right)
\bigg] \nn \\
&&
+\ \pi^2\frac{p_{\mu}p_{\nu}}{p^2}\bigg[\frac{m_{\pm}}{2} - \bigg(\frac{3p^2 +4m_{\pm}^2}8\bigg)
I\left(\frac{p}{2},m_\pm \right)
\bigg]\;. \label{result.foton}
\eea
Using the results~(\ref{K})-(\ref{result.foton}) into the decomposition~(\ref{foton.basis}) 
and making the replacement $m_{\pm} \to m$ we obtain the photon propagator in accordance to the results in Refs.~\cite{Delbourgo, Chaichian}. This completes our calculation of the one-loop photon propagator. Next we calculate the fermion propagator at one-loop.

\section{ONE-LOOP FERMION PROPAGATOR}

\begin{figure}[t]
\begin{center}
\includegraphics[width=0.25\textwidth]{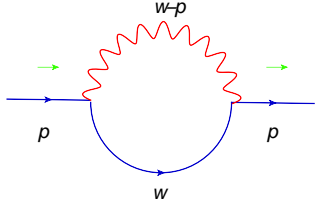}
\end{center}
\caption{One-loop correction to the fermion propagator.}
\label{fig2}
\end{figure}

The one-loop correction to the fermion propagator has been previously considered by some of us in Ref.~\cite{raya}, and corresponds to the diagram in Fig.~\ref{fig2}. To calculate this correction, we write the one-loop fermion propagator as
\begin{equation}
S_F^{(1)}(p)= S_+^{(1)}(p)+S_-^{(1)}(p)\;,
\end{equation}
where
\begin{equation}
S_\pm(p)=-\frac{A_\pm^{(1)}(p){\not \! p}+B_\pm^{(1)}(p)}{\left[ A_\pm^{(1)}(p)\right]^2 p^2+\left[B_\pm^{(1)}(p)\right]^2}\chi_\pm\;.
\label{1lp}
\end{equation}
The functions $A_\pm^{(1)}(p)$ and $B_\pm^{(1)}(p)$ are thus obtained as
\begin{eqnarray}
A_\pm^{(1)}(p) &=& -\frac{2\pi\alpha}{p^2} \int \frac{d^3w}{(2\pi)^3}{\rm Tr}\left[{\not \! p} \gamma_\mu S_\pm^{(1)}(w) \gamma_\nu  \Delta_{\mu\nu}^{(0)}(w-p;\xi)\chi_\pm \right]\;, \nonumber\\
B_\pm^{(1)}(p) &=& 2\pi\alpha \int \frac{d^3w}{(2\pi)^3}{\rm Tr}\left[\gamma_\mu S_\pm^{(1)}(w) \gamma_\nu  \Delta_{\mu\nu}^{(0)}(w-p;\xi)\chi_\pm \right]\;.
\end{eqnarray}
Performing the traces, after suitable manipulations, these functions can be expressed in the form (compare against Eq~(26) of Ref.~\cite{raya})
\begin{eqnarray}
A_\pm^{(1)}(p) &=& -\frac{\alpha\xi}{4\pi^2}\frac{(p^2\!-\!m^2)^2}{p^2}\left[{\cal I}_2(p^2,m^2)\!-\!{\cal I}(0,m^2_\pm,p^2) \right]
\nonumber\\
&&+\frac{\alpha}{4\pi^2\theta^2p^2}\Bigg[ (p^2-m^2)^2{\cal I}(0,m^2_\pm,p^2)-\left[(p^2-m^2)^2+\theta^4 \right] {\cal I}(\theta^2,m^2_\pm,p^2)- \theta^2{\cal T}(m^2)\Bigg]
\nonumber\\
&&\mp \frac{\alpha m_\pm}{2\pi^2p^2}\Bigg[ (p^2+m^2-\theta^2){\cal I}(\theta^2,m^2_\pm,p^2)-(p^2+m^2){\cal I}(0,m^2,p^2)-{\cal T}(\theta^2)\Bigg]  
\;,\nonumber\\
B_\pm^{(1)}(p) &=& \frac{\alpha \xi m_\pm}{2\pi^2}{\cal I}(0,m^2_\pm,p^2) + \frac{\alpha m_\pm}{\pi^2} {\cal I}(\theta^2,m_\pm^2,p^2) \nonumber\\
&&\mp \frac{\alpha}{2\pi^2\theta}\Bigg[ (p^2+\theta^2+m_\pm^2){\cal I}(\theta^2,m_\pm^2,p^2)-(p^2+m_\pm^2){\cal I}(0,m^2,p^2)-{\cal T}(\theta^2)\Bigg]\;,
\label{decompSf}
\end{eqnarray}
where the master integrals are
\begin{eqnarray}
&&\hspace{-5mm}
{\cal T}(m^2_\pm)\ =\ \int d^3w \frac{1}{w^2+m^2_\pm}\;,\nonumber\\
&&\hspace{-8mm}
{\cal I}(\theta^2,m^2_\pm,p^2)\ =\ \int d^3w \frac{1}{[w^2+\theta^2][(p-w)^2+m_\pm^2]},\nonumber\\
&&\hspace{-5mm}
{\cal I}_2(m^2_\pm,p^2)\ =\ \int d^3w \frac{1}{w^4[(p-w)^2+m_\pm^2]}.\label{masterfp}
\end{eqnarray}
These integrals are performed using standard techniques (see Appendix), and give
%. As they appear several times in the evaluation of other master integrals, we quote their results here
\begin{eqnarray}
{\cal T}(m^2_\pm)&=& -2\pi^2m_\pm\;,\nonumber\\
{\cal I}(\theta^2,m^2_\pm,p^2)&=& 2\pi^2 I(p,m_\pm^2+\theta^2)\;,\nonumber\\
{\cal I}_2(m^2_\pm,p^2)&=&-\frac{2\pi^2m_\pm}{(p^2+m_\pm^2)^2}\;.
\label{res}
\end{eqnarray}
Inserting the above results into the decompositions~(\ref{decompSf}) and~(\ref{1lp}), we obtain the one-loop correction to the fermion propagator, in accordance with Eq.~(27) of Ref.~\cite{raya}. This completes the one-loop correction of the two-point Green's functions corresponding to the Lagrangian~(\ref{lag}). Next we calculate the one-loop correction to the fermion-photon vertex.

\section{ONE-LOOP FERMION-PHOTON VERTEX}

\begin{figure}[b]
\begin{center}
\includegraphics[width=0.25\textwidth]{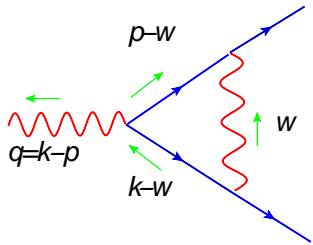}
\end{center}
\caption{One-loop correction to the fermion-photon vertex.}
\label{fig3}
\end{figure}

At ${\cal O}(\alpha)$, the fermion-photon vertex is depicted in Fig.~\ref{fig3}, and corresponds to 
\begin{eqnarray}
\Gamma_{\mu}(k,p)=\,\gamma_{\mu}+\,\Lambda_{\mu} \;.  \label{01loopvertex}
\end{eqnarray}
Using the Feynman rules, $\Lambda_{\mu}$  is simply given by
\begin{eqnarray}
\Lambda_{\mu}=
 e^2 \int\frac{d^3w}{(2\,\pi)^3}
\gamma_{\alpha}{\it S}^{(0)}_{F}(p-w)\gamma_{\mu}
{\it S}^{(0)}_{F}(k-w)\gamma_{\beta}
\Delta^{(0)}_{\alpha\beta}(w). \nn \\  \label{1loopvertex}
\end{eqnarray}
Given the properties of the projectors in (\ref{properties projectors}), the crossed structures vanish and therefore we have only integrals with the same mass propagators ($m_{+}$ or $m_{-}$). Given this, we can write $\Lambda_{\mu}$ as
\be
\Lambda_{\mu}\equiv \Lambda_{(+)\ \mu} + \Lambda_{(-)\ \mu}
\ee
where
\begin{eqnarray}
\Lambda_{(\pm)\ \mu}&=&\frac{\,{\alpha}}{2\,{\pi}^2}
\{ \left[ \gamma_{\sigma} {\not \! p}\,{\gamma_{\mu}}\,{\not \! k} 
 \gamma_{\sigma}   + m_{\pm} ( 4 k_{\mu} + 4 p_{\mu} -  {\not \! p} 
{\gamma_{\mu}} -  {\gamma_{\mu}}   {\not \! k} ) -m_{\pm}^2
{\gamma_{\mu}}       \right] 
{\it J}^{(0)}(\theta,m_{\pm})  \nonumber \\
&&
\hspace{10mm} -  \left[ {\gamma_{\sigma}}
 {\not \! p}\,{\gamma_{\mu}}{\gamma_{\nu}} \gamma_{\sigma}
+ \gamma_{\sigma}  {\gamma_{\nu}} {\gamma_{\mu}} {\not \! k} \gamma_{\sigma}
- 6 m_{\pm} \delta_{\mu \nu}     \right] 
{{\it J}_{\nu}^{(1)}}(\theta,m_{\pm})
+{\gamma_{\sigma}}{\gamma_{\nu}}{\gamma_{\mu}}{\gamma_{\lambda}}
\gamma_{\sigma}{{\it J}_{\nu\lambda}^{(2)}}(\theta,m_{\pm})\nonumber\\
&& \hspace{10mm}
+\xi[{\gamma_{\mu}}{\it K}^{(0)}(m_{\pm}) - 
\left[ {\gamma_{\nu}}{\not \! p}\,{\gamma_{\mu}}+\,
{\gamma_{\mu}}{\not \! k}\,{\gamma_{\nu}}-2m_{\pm} \delta_{\mu \nu}
\right]
{\it J}_{\nu}^{(1)}(m_{\pm})
\nonumber\\
&& \hspace{10mm} +  \left[ {\gamma_{\nu}}{\not \! p}\,
{\gamma_{\mu}}{\not \! k}\,
{\gamma_{\lambda}}
+m_{\pm}(\,{\gamma_{\nu}}{\not \! p}\,{\gamma_{\mu}}\,{\gamma_{\lambda}}+
\,{\gamma_{\nu}}\,{\gamma_{\mu}}\,{\not \! k} \,{\gamma_{\lambda}})
+m^2_{\pm}\,{\gamma_{\nu}}\,{\gamma_{\mu}}\,{\gamma_{\lambda}}  \right] 
{{\it I}_{\nu\lambda}^{(2)}}(m_{\pm})]
\nonumber\\
&& \hspace{10mm}
- {\gamma_{\mu}}[{\it K}^{(0)}(m_{\pm}) -\theta^2{\it J}^{(0)}(\theta,m_{\pm}) ] - 
\left[ {\gamma_{\nu}}{\not \! p}\,{\gamma_{\mu}}+\,
{\gamma_{\mu}}{\not \! k}\,{\gamma_{\nu}}-2m_{\pm} \delta_{\mu \nu}
\right]
{\it J}_{\nu}^{(1)}(\theta, m_{\pm})
\nonumber\\
&& \hspace{10mm} +  \left[ {\gamma_{\nu}}{\not \! p}\,
{\gamma_{\mu}}{\not \! k}\,
{\gamma_{\lambda}}
+m_{\pm}(\,{\gamma_{\nu}}{\not \! p}\,{\gamma_{\mu}}\,{\gamma_{\lambda}}+
\,{\gamma_{\nu}}\,{\gamma_{\mu}}\,{\not \! k} \,{\gamma_{\lambda}})
+m^2_{\pm}\,{\gamma_{\nu}}\,{\gamma_{\mu}}\,{\gamma_{\lambda}}  \right] 
{{\it L}_{\nu\lambda}^{(2)}}(\theta,m_{\pm})
\nonumber\\
&&\hspace{10mm}
-\theta\varepsilon_{\nu\lambda\rho}[-  
\left[ {\gamma_{\nu}}{\not \! p}\,{\gamma_{\mu}}\,{\gamma_{\rho}}\,{\gamma_{\lambda}}+\,
{\gamma_{\nu}}\,{\gamma_{\rho}}\,{\gamma_{\mu}}\,{\not \! k}\,{\gamma_{\lambda}}+m_{\pm}(\gamma_{\nu}\,{\gamma_{\rho}}\,{\gamma_{\mu}}\,{\gamma_{\lambda}} + \gamma_{\nu}\,{\gamma_{\mu}}\,{\gamma_{\rho}}\,{\gamma_{\lambda}}) 
\right]
{\it J}^{(0)}(\theta, m_{\pm})
\nonumber\\
&& \hspace{10mm} + \gamma_{\nu}\,{\gamma_{\sigma}}\,{\gamma_{\mu}}\,\gamma_{\rho}\,{\gamma_{\lambda}}{\it J}^{(1)}_{\sigma}(\theta, m_{\pm}) + [ {\gamma_{\nu}}{\not \! p}\,
{\gamma_{\mu}}{\not \! k}\,
{\gamma_{\lambda}}
+m_{\pm}(\,{\gamma_{\nu}}{\not \! p}\,{\gamma_{\mu}}\,{\gamma_{\lambda}}+
\,{\gamma_{\nu}}\,{\gamma_{\mu}}\,{\not \! k} \,{\gamma_{\lambda}})
\nn\\
&& \hspace{10mm}
+ m^2_{\pm}\,{\gamma_{\nu}}\,{\gamma_{\mu}}\,{\gamma_{\lambda}} ] 
{{\it L}_{\rho}^{(1)}}(\theta,m_{\pm})]
\}\chi_{\pm}. \label{1loopvertexevaluated}
\end{eqnarray}

In this expression we have introduced the basis integrals
${K^{(0)}}$, $J^{(0)}$, $J^{(1)}_{\mu}$, $J^{(2)}_{\mu\nu}$, $I^{(0)}$, 
$I^{(1)}_{\mu}$, $I^{(2)}_{\mu\nu}$, $L^{(1)}_{\mu}$, $L^{(2)}_{\mu\nu}$ which we define as
\begin{eqnarray}
{\it K}^{(0)}({m_{\pm}})&=&\int\,d^3w\,\frac{1}{[(p-w)^2+m^2_{\pm}]\,[(k-w)^2+m^2_{\pm}]}\;, 
\nonumber \\
{\it J}^{(0)}(m_{\pm})&=&\int\,d^3w\,\frac{1}{w^2\,[(p-w)^2+m^2_{\pm}]\,[(k-w)^2+m^2_{\pm}]}\;,
\nonumber \\
{\it J}^{(0)}(\theta, m_{\pm})&=&
\int\,d^3w\,\frac{1}{[w^2 + \theta^2]\,[(p-w)^2+m^2_{\pm}]\,[(k-w)^2+m^2_{\pm}]}\;,\nn
%\eea
\nonumber \\
%
%
%
%\bea
{\it J}^{(1)}_{\mu}(m_{\pm})&=&
\int\,d^3w\,
\frac{w_{\mu}}{w^2\,[(p-w)^2+m^2_{\pm}]\,[(k-w)^2+m^2_{\pm}]}\;,
\nonumber \\
{\it J}^{(1)}_{\mu}(\theta, m_{\pm})&=&
\int\,d^3w\,
\frac{w_{\mu}}{[w^2 + \theta^2]\,[(p-w)^2+m^2_{\pm}]\,[(k-w)^2+m^2_{\pm}]}\;,
\nonumber \\
{\it J}^{(2)}_{\mu\nu}( m_{\pm})&=&
\int\,d^3w\,
\frac{w_{\mu}w_{\nu}}{w^2\,[(p-w)^2+m^{2}_{\pm}]\,[(k-w)^2+m^{2}_{\pm}]}\;,
\nonumber \\
{\it J}^{(2)}_{\mu\nu}(\theta, m_{\pm})&=&
\int\,d^3w\,
\frac{w_{\mu}w_{\nu}}{[w^2 + \theta^2]\,[(p-w)^2+m^{2}_{\pm}]\,[(k-w)^2+m^{2}_{\pm}]}\;, \nn
\\%\eea
%\nonumber \\
%
%
%
%\bea
{\it I}^{(0)}(m_{\pm})&=&
\int\,d^3w\,
\frac{1}{w^4\,[(p-w)^2+m^2_{\pm}]\,[(k-w)^2+m^2_{\pm}]}\;,
\nonumber \\
{\it I}^{(1)}_{\mu}(m_{\pm})&=&
\int\,d^3w\,
\frac{w_{\mu}}{w^4\,[(p-w)^2+m^2_{\pm}]\,[(k-w)^2+m^2_{\pm}]}\;,
\nonumber \\
{\it I}^{(2)}_{\mu\nu}(m_{\pm})&=&
\int\,d^3w\,
\frac{w_{\mu}w_{\nu}}{w^4\,[(p-w)^2+m^2_{\pm}]\,[(k-w)^2+m^2_{\pm}]}\;,
\nonumber \\
{\it L}^{(1)}_{\mu}(\theta,m_{\pm})&=&
\int\,d^3w\,
\frac{w_{\mu}}{w^2\,[w^2 +\theta^2]\,[(p-w)^2+m^2_{\pm}]\,[(k-w)^2+m^2_{\pm}]}\;,
\nonumber \eea
\bea
{\it L}^{(2)}_{\mu\nu}(\theta,m_{\pm})&=&\
\int\,d^3w\,
\frac{w_{\mu}w_{\nu}}{w^2\,[w^2 +\theta^2]\,[(p-w)^2+m^2_{\pm}]\,[(k-w)^2+m^2_{\pm}]}
\label{integrals}  \;.
\end{eqnarray}

From this list of integrals, those which depend only on fermion masses have been calculated in~\cite{Raya-Bashir} and for them we just replace $m\to m_{\pm}$ to obtain the results needed in the calculation at hand. The rest correspond to two-, three- and four-point integrals that depend both on
a fermion and a gauge boson mass, therefore they might be useful in theories in other space-time dimensions. 
We solve these in arbitrary dimensions in the Appendix. Also, it is important to mention that these results are crucial to our interest in the impact of the CS term on certain physical observables related to the Lagrangian~(\ref{lag}).

These results along with the decomposition proposed in Eq.~(\ref{1loopvertexevaluated}) complete the calculation of the one-loop correction of the fermion-photon vertex. We summarize our findings in the next section.

\section{Summary and conclusions}

We have derived the one-loop correction to the photon propagator, fermion propagator and the fermion-photon vertex in Maxwell-Chern-Simons in QED$_{3}$. In doing so, we have proposed a basis of master integrals to expand said Green functions, some of them new in literature. Due to the potential applications of two- and three-point Green functions with a massive gauge boson in physical processes in several frameworks of particle and condensed matter physics, we evaluated these new integrals in arbitrary space-time dimensions. The basis expansion contains the ordinary expansion of QED$_3$ as the particular case of our findings in the parity preserving limit~\cite{Raya-Bashir}, which is achieved when both the CS term and the Haldane mass are taken to zero. Massless theory for both fermions and photons~\cite{bashir} follows as well. Green functions we have obtained will help us make progress in certain physical applications in the near future. Particularly, we are interested in the radiative generation of fermion masses by the CS term and its impact in several physical observables. This project is in progress and the results will be presented elsewhere.

\begin{acknowledgements}
We acknowledge SNI, CIC-UMSNH and CONACyT for support. We also acknowledge Adnan Bashir for clarifying discussions and bringing to our attention important references. Y.C.S. acknowledges support from CONACyT and Universidad de Sonora.
\end{acknowledgements}

\section*{Appendix}

In this Appendix we provide details in the calculations for three sets of master integrals:
for the one-loop photon propagator described in (\ref{result.foton}), the one-loop fermion propagator, Eq.~(\ref{masterfp}), and for the one-loop vertex 
described in (\ref{integrals}).
Let us begin with the $K$-type master integrals. \\

\begin{center} 
{\bf 1. {\bf $K$}} 
\end{center}

We begin solving $K$. Using   
\be
\frac{1}{ab}=\int_{0}^{1}\frac{dx}{[xa + (1 - x)b]^{2}} \;,
\ee 
%where
%\be 
%a=(w-p)^{2} + m^2_{\pm}\;,\qquad b=w^{2} + m^{2}_{\pm}\;. 
%\ee 
we can write
\bea
K&=&\int_{0}^{1}dx \int d^{3}w
\frac{1}{\bigg[x[(w -p)^{2} +
m^{2}_{\pm}] + (1 - x)(w^{2} + m^{2}_{\pm})\bigg]^{2}} \;. 
\eea
The denominator in the above expression can be written as 
%\bea
%{\rm denominator}&=&%x[(w -p)^{2} + m^{2}_{\pm}] + (1 - x)(w^{2} + m^{2}_{\pm})] \nn\\
%&=& x[w^{2} - 2w\cdot p + p^{2} + m^{2}_{\pm}] + (1 - x)w^{2} + m^{2}_{\pm}(1 -x) \nn\\ &=& 
$w^{2} - 2xw\cdot p + xp^{2} + m^{2}_{\pm}$,
%\eea 
which after the change of variables $w'=w - px$, allows us to write
\be 
K=\int_{0}^{1}dx \int d^{3}w'\frac{1}{[(w')^{2} + p^{2}x(1 - x) + m^{2}_{\pm}]^{2}} \;. 
\ee 
This integral has the following form: 
\be
\int\frac{d^Dw}{(w^2 - s)^n}=(-1)^n\pi^{D/2}\frac{\Gamma\big(n - \frac{D}{2}\big)}{\Gamma(n)}\frac{1}{s^{n - D/2}}\;. 
\label{dimreg1}
\ee
In our case, we take  $D=3$, $n=2$ and $s=-p^{2}x(1 - x) - m^{2}_{\pm}$, and then 
\bea
K=\pi^{3/2}\Gamma\bigg(\frac{1}{2}\bigg)\int_{0}^{1}dx[- p^{2}x(1 - x)
- m^{2}_{\pm}]^{-1/2}=\frac{2\pi^2}{p}\arctan\bigg(\frac{p}{2m_{\pm}}\bigg)\;.
\eea 
%in agreement with \cite{raya} and \cite{bashir}. 
\begin{center}
{\bf 2. {\bf $K_{\mu}$}} 
\end{center}

Now, we calculate $K_{\mu}$.
%\bea
%K_{\mu}=\int_{E}d^3w\frac{w_{\mu}}{[w^2 + m_{\pm}^2][(w -p)^2+ m_{\pm}^2]}
%\eea
This integral depends only of $p_{\mu}$, thus we can write    
\bea 
K_\mu\equiv cp_{\mu} \;. 
\eea 

We find $c$ multiplying with $p_{\mu}$ on both sides 
\be 
cp^{2}=\int d^{3}w\frac{w \cdot p}{(w^{2} + m^{2}_{\pm})[(w -
p)^{2} + m^{2}_{\pm}]} \;. 
\ee 
Using $w \cdot p=w^{2} + p^{2} - (w - p)^{2}$, we have
\bea 
c&=&%\frac{1}{2p^{2}}\int_{E}d^{3}w\frac{w^{2} + p^{2} - (w - p)^{2}}{(w^{2} + m^{2}_{\pm})[(w - p)^{2}+ m^{2}_{\pm}]} \nn\\
%&=&\frac{1}{2}\int_{E} \frac{d^{3}w}{(w^{2} + m^{2}_{\pm})[(w - p)^{2}+ m^{2}_{\pm}]} 
\frac{1}{2}K\;. 
\eea 
%Thus, 
%\be K_\mu\int_{E} d^{3}w \frac{w^{\mu}}{(w^{2} +
%m^{2}_{\pm})[(w - p)^{2} + m^{2}_{\pm}]}=\frac{1}{2}\int_{E}
%d^{3}w\frac{p^{\mu}}{(w^{2} + m^{2}_{\pm})[(w - p)^{2} + m^{2}_{\pm}]} \;, \ee

\begin{center}
{\bf 3. {\bf ${K_{\mu\nu}}$}} 
\end{center}

Finally, we solve $K_{\mu \nu}$.  
%\bea
%K_{\mu \nu}&=&\int_{E}d^3w\frac{w_{\mu}w_{\nu}}{[w^2 + m_{\pm}^2][(w -p)^2+ m_{\pm}^2]}\;.
%\eea
%We write $K^{\mu \nu}$ 
In its most general form, it depends only on $p_{\mu}p_{\nu}$ and $\delta_{\mu\nu}$,
\be 
K_{\mu \nu}
%=\int_{E}d^3w\frac{w^{\mu}w^{\nu}}{[w^{2} + m^{2}_{\pm}][(w - p)^{2} +m^{2}_{\pm}]}
\equiv a\bigg[\frac{p_{\mu}p_{\nu}}{p^{2}} +
\frac{1}{3}\delta_{\mu \nu} \bigg] - b\delta_{\mu \nu}\;, 
\ee
which is symmetrical under the exchange of $\mu$ and $\nu$. After simplifying, we find that
\bea 
a=\frac{1}{4}\int d^{3}w\frac{1}{w^{2} + m^{2}_{\pm}} - \frac{3p^{2} +
4m^{2}_{\pm}}{8} K\;, 
%\int_{E} d^{3}w\frac{1}{[w^{2} + m^{2}_{\pm}][(w - p)^{2} +m^{2}_{\pm}]}  
\eea
and
\be 
b=\frac{1}{3}\int \frac{d^{3}w}{(w
-p)^{2} + m^{2}_{\pm}} + \frac{m^{2}_{\pm}}{3}K\;, 
%\int_{E} \frac{d^{3}w}{[w^{2} +m^{2}_{\pm}][(w - p)^{2} + m^{2}_{\pm}]}  
\ee
where the tadpole integrals [see Eq.~(\ref{tadpole}) below] are
\bea
{\cal T}(m_\pm^2)\ =\ \int d^{3}w\frac{1}{w^{2} + m^{2}_{\pm}}= \int \frac{d^{3}w}{(w
-p)^{2} + m^{2}_{\pm}}= -2\pi^{2}m_{\pm}\;. 
\eea
%Then
%\bea
%a=-\frac{\pi^2 m_{\pm}}{2} - \frac{(3p^2 + 4m^2_{\pm})}{4}\frac{\pi^2}{p}\arctan\left(\frac{p}{2m_{\pm}}\right)
%\eea
%and
%\bea
%b=-\frac{2}{3}\pi^2m_{\pm} + \frac{2\pi^2m_{\pm}^2}{3p}\arctan\left(\frac{p}{2m_{\pm}}\right)
%\eea
%Finally, $K^{\mu \nu}$ is
%\bea
%K^{\mu \nu}&=&\delta_{\mu \nu}\bigg[\frac{\pi^2m_{\pm}}{2} - \bigg(\frac{p^2 + 4m_{\pm}^2}{4}\bigg)\frac{\pi^2}{p}\arctan\left(\frac{p}{2m_{\pm}}\right)\bigg] \nn \\
%&&- \frac{p_{\mu}p_{\nu}}{p^2}\bigg[\frac{\pi^2m_{\pm}}{2} + \bigg(\frac{3p^2 +4m_{\pm}^2}{4}\bigg)\frac{\pi^2}{p}\arctan\left(\frac{p}{2m_{\pm}}\right)\bigg]\;.
%\eea
Next, we provide details of the calculation of the integrals in Eq.~(\ref{masterfp}).\\

\begin{center}
{\bf 4. {\bf ${\cal T}(m_{\pm}^2)$}}
\end{center}

The integral
\begin{equation}
{\cal T}(m_\pm^2)=\int d^3w \frac{1}{w^2+m_\pm^2}\label{tadpole}
\end{equation}
can be evaluated directly from Eq.~(\ref{dimreg1}) with $n=1$, $D=3$ and $s=-m_\pm^2$. It yields
\be
{\cal T}(m_\pm^2)=-2\pi^2m_\pm.
\ee
\newpage
\begin{center}
{\bf 5. {\bf ${\cal I}(\theta^2,m_{\pm}^2,p^2)$}}
\end{center}

Let us evaluate
\be
{\cal I}(\theta^2,m_\pm^2,p^2)=\int d^3w \frac{1}{[w^2+\theta^2][(p-w)^2+m_\pm^2]}.
\ee
Using Feynman parametrization, and after the appropriate change of variables, we can write
\bea
{\cal I}(\theta^2,m_\pm^2,p^2)&=&
\int_0^1 \int d^3w' \frac{1}{[(w')^2+x(1-x)p^2+m_\pm^2x+\theta^2(1-x)]^2}\nonumber\\
&=& \pi^2\int_0^1 dx (x(1-x)p^2+m_\pm^2x+\theta^2(1-x))^{-\frac{1}{2}}\nonumber\\
&=&
2\pi^2I(p,m_\pm^2+\theta^2)\;.
\eea

\begin{center}
{\bf 6. {\bf ${\cal I}_2(m_{\pm}^2,p^2)$}}
\end{center}

In order to evaluate
\be
{\cal I}_2(m_\pm^2,p^2)=\int d^3w \frac{1}{w^4[(p-w)^2+m_\pm^2]}\;,
\ee
we introduce the auxiliary integral
\be 
{\cal I}_\mu(m_\pm^2,p^2)=\int d^3w \frac{w_\mu}{w^4[(p-w)^2+m_\pm^2]}\;,
\ee
such that
\bea
p_\mu {\cal I}_\mu(m_\pm^2,p^2)=\frac{1}{2}((p^2+m_\pm^2){\cal I}_2(m_\pm^2,p^2)+{\cal I}(0,m_\pm^2,p^2))\;.
\eea
Using the identity
\bea
\frac{1}{A^nB^p}=\frac{\Gamma(n+p)}{\Gamma(n)\Gamma(p)} 
\int_0^1 dx x^{n-1}(1-x)^{p-1} \frac{1}{[x A + (1-x) B]^{n+p}}\;,
\eea
with $n=1$, $p=2$, followed by the shift $w'=w-px$, we can write
\bea
{\cal I}_\mu(m_\pm^2,p^2)=2\int_0^1dx(1-x)\int d^3w' \frac{p_\mu x}{[(w')^2+p^2x(1-x)+m_\pm^2x]^3}\;,
\eea
performing the $w'$ integration,
\bea
{\cal I}_\mu(m_\pm^2,p^2)=\frac{\pi^2}{2}p_\mu \int_0^1 dx (x^{-\frac{1}{2}}-x^\frac{1}{2})[p^2(1-x)+m_\pm^2]^{-\frac{3}{2}}\;,
\eea
and the remaining integration yields
\bea
{\cal I}_\mu(m_\pm^2,p^2)&=&\frac{\pi^2}{p^2}p_\mu\left[-\frac{m_\pm}{p^2+m_\pm^2}+I(p,m_\pm) \right]\;.
\eea
Thus
\bea
{\cal I}_2(m_\pm^2,p^2)= \frac{2p_\mu {\cal I}_\mu(m_\pm^2,p^2)-{\cal I}(0,m_\pm^2,p^2)}{p^2+m_\pm^2}=-\frac{2\pi^2m_\pm}{(p^2+m_\pm^2)^2}\;.
\eea
%\end{document}

Next, we will give the results for the new integrals that are described in~(\ref{integrals}). \\

\begin{center}
{\bf 7. {\bf $J^{(0)}(\theta,m_{\pm})$}}
\end{center}
We begin solving $J^{(0)}(\theta, m_{\pm})$.
For this purpose, we consider the following integral in Minkowski space:
\be
{\cal J}^{(0)}_{\nu_1,\nu_2,\nu_3}(\theta, m_{\pm})=
\int_M d^D w\frac{1}{[(p - w)^2 - m_{\pm}^2]^{\nu_{1}}[w^2 - \theta^2]^{\nu_{2}}[(k - w)^2 - m_{\pm}^2]^{\nu_{3}}}\;. 
\ee
Using Schwinger parametrization
\be
\frac{1}{[A_1+A_2]^\nu} = \frac{1}{2\pi i}\int_{-i\infty}^{i\infty} d\zeta A_1^\zeta A_2^{-\nu-\zeta}\frac{\Gamma(-\zeta) \Gamma(\nu+\zeta)}{\Gamma(\nu)}
\label{schwingerparam}
\ee
to each of the factors in the denominator, the Mellin-Barnes representation for ${\cal J}^{(0)}_{\nu_1,\nu_2,\nu_3}(\theta, m_{\pm})$ becomes
\bea
{\cal J}^{(0)}_{\nu_1,\nu_2,\nu_3}(\theta, m_{\pm})&\equiv&\frac{1}{(2\pi
i)^3\Gamma(\nu_{1})\Gamma(\nu_{2})\Gamma(\nu_{3})}\int_{-i\infty}^{+i\infty}\int_{-i\infty}^{+i\infty}\int_{-i\infty}^{+i\infty}d\lambda_{3}d\lambda_{34}d\lambda_{5}(-m^2)^{\lambda_{34}}(-\theta^2)^{\lambda_{5}} \nn \\ && \times\Gamma(-\lambda_{3})\Gamma(-\lambda_{34} + \lambda_{3})\Gamma(-\lambda_{5})\Gamma(\lambda_{3}
+ \nu_{1}) \Gamma(\lambda_{34} -\lambda_{3} + \nu_{3})\Gamma(\lambda_{5} + \nu_{2})\nn \\ &&\times{\cal J}^{(0)}_{\nu_{1}+ \lambda_{3},\nu_{2} + \lambda_{5} ,\nu_{3}+ \lambda_{34} - \lambda_{3}}(0,0)\;, 
\eea
where we have used the notation $\lambda_{34}=\lambda_{3} + \lambda_{4}$ and $d\lambda_{4}=d\lambda_{34}\;,$ and 
where $J^{(0)}_{\nu_{1}+ \lambda_{3},\nu_{2}+ \lambda_{5} ,\nu_{3}+ \lambda_{34} - \lambda_{3}}(0,0)$  has been calculated in Ref.~\cite{davydychev}. Replacing these results  and simplifying, we obtain the following expression in terms of five integrals to evaluate
\bea
{\cal J}^{(0)}_{\nu_1,\nu_2,\nu_3}(\theta, m_{\pm})&=&\frac{(i)^D}{\prod_{i=1}^{3}\Gamma(\nu_{i})}\frac{(k^2)^{D/2
-\nu_{123}}}{(2\pi
i)^5}\int\int\int\int\int_{-i\infty}^{+i\infty}d\lambda_{1}d\lambda_{2}d\lambda_{3}d\lambda_{34}d\lambda_{5}
\Gamma(-\lambda_{1})
\nn \\ && \times\Gamma(-\lambda_{2})\Gamma(-\lambda_{3})
\Gamma(\lambda_{3}-\lambda_{34})\Gamma(-\lambda_{5})\bigg(\frac{p^2}{k^2}\bigg)^{\lambda_{1}}\bigg(\frac{q^2}{k^2}\bigg)^{\lambda_{2}}
\bigg(-\frac{m^2}{k^2}\bigg)^{\lambda_{34}}\bigg(-\frac{\theta^2}{k^2}\bigg)^{\lambda_{5}} 
\nn \\ &&\times \frac{\Gamma\big(\lambda_{12}
+ \lambda_{34} + \nu_{123} + \lambda_{5}- \frac{D}{2}\big)}{\Gamma(D - \nu_{123} -\lambda_{34}- \lambda_{5})}
\Gamma(\lambda_{123} +
\nu_{1})\Gamma\bigg(-\lambda_{13}-\nu_{12} - \lambda_{5}
+\frac{D}{2}\bigg)\nn \\ &&\times\Gamma\bigg(-\lambda_{2} -\lambda_{34} -\nu_{13} +
\frac{D}{2}\bigg)\;. \label{J0}
\eea 
We perform one integration at a time. First we integrate over $\lambda_{3}$, defining
\bea I_{\lambda_{3}}=\frac{1}{2\pi
i}\int_{-i\infty}^{+i\infty}d\lambda_{3}\Gamma(-\lambda_{3})\Gamma(\lambda_{3}
- \lambda_{34})\Gamma(\lambda_{123} +
\nu_{1})\Gamma\bigg(-\lambda_{13} - \nu_{12} -\lambda_{5}+ \frac{D}{2}\bigg) \;. 
\eea 

\begin{figure}[t]
\begin{center}
\includegraphics[width=0.45\textwidth]{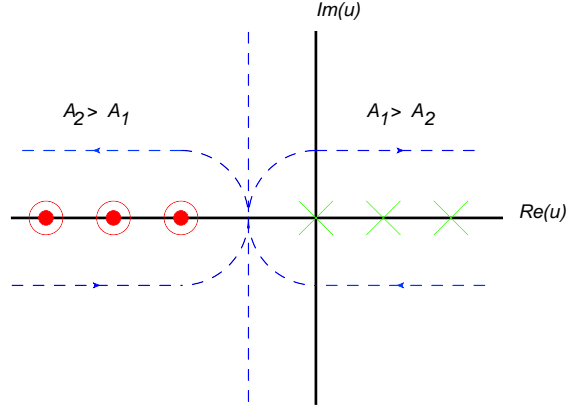}
\end{center}
\caption{Mellin-Barnes contour of integration which separates the poles of $\Gamma(\ldots+u)$ from those of $\Gamma(\ldots-u)$. We close the contour to the left or to the right, choosing one of the two series of poles. $A_1$ and $A_2$ represent the arguments in the sum which are given a Schwinger representation~(\ref{schwingerparam}).}
\label{fig4}
\end{figure}

Using the Barnes lemma 
\bea \frac{1}{2\pi
i}\int_{-i\infty}^{+i\infty}ds\Gamma(a + s)\Gamma(b + s)\Gamma(c -
s)\Gamma(d - s)=\frac{\Gamma(a + c)\Gamma(b + c)\Gamma(a +
d)\Gamma(b + d)}{\Gamma(a + b + c + d)}\;,
\eea 
with 
$a=-\lambda_{34}\;, b=\nu_{1} + \lambda_{12}\;, c=0\;, d= \frac{D}{2} - \nu_{12} - \lambda_{1} - \lambda_{5}\;,$
we find that
\bea
I_{\lambda_{3}}&=&\frac{\Gamma(-\lambda_{34})\Gamma(\nu_{1} +\lambda_{12})
\Gamma\big(\lambda_{2}-\nu_{2}- \lambda_{5} + \frac{D}{2}\big)}{\Gamma\big(\lambda_{2} -\lambda_{34}
-\nu_{2} -\lambda_{5}+ \frac{D}{2}\big)}\Gamma\big(-\lambda_{1}-\lambda_{34}-\lambda_{5}-\nu_{12}+\frac{D}{2}\big)\;. \nn
\eea 
Then, after integrating over $\lambda_{3}$, Eq.~(\ref{J0}) becomes 
\bea
{\cal J}^{(0)}_{\nu_1,\nu_2,\nu_3}(\theta, m_{\pm})&=&\frac{(i)^D}{\prod_{i=1}^{3}\Gamma(\nu_{i})}\frac{(k^2)^{D/2
-\nu_{123}}}{(2\pi
i)^4}\int\int\int\int_{-i\infty}^{+i\infty}d\lambda_{1}d\lambda_{2}d\lambda_{34}d\lambda_{5}\Gamma(-\lambda_{1})
\Gamma(-\lambda_{2})\nn \\ && \times\Gamma(-\lambda_{5})\bigg(\frac{p^2}{k^2}\bigg)^{\lambda_{1}}\bigg(\frac{q^2}{k^2}\bigg)^{\lambda_{2}}
\bigg(-\frac{m^2}{k^2}\bigg)^{\lambda_{34}}\bigg(-\frac{\theta^2}{k^2}\bigg)^{\lambda_{5}}\nn \\ &&\times
\frac{\Gamma\big(\lambda_{12}
+ \lambda_{34} + \nu_{123} + \lambda_{5} - \frac{D}{2}\big)\Gamma\big(-\lambda_{2}
- \lambda_{34} -\nu_{13} + \frac{D}{2}\big)}{\Gamma(D -\nu_{123}
-\lambda_{34} -\lambda_{5})}\nn \\ &&\times
\frac{\Gamma(-\lambda_{34})\Gamma(\nu_{1} +
\lambda_{2})\Gamma\big(-\lambda_{1}-\lambda_{34} -\lambda_{5}-\nu_{12}+\frac{D}{2}
\big)}{\Gamma\big(\lambda_{2} -\lambda_{34} -\nu_{2} -\lambda_{5}+
\frac{D}{2}\big)}\nn \\ && \times  \Gamma\big(\lambda_{2} -\nu_{2} -\lambda_{5}+
\frac{D}{2}\big)\;. 
\eea 
One particular limit of our interest is $\theta\to 0$. We integrate over $\lambda_5$ first and call $I_{\lambda_{5}}$ the relevant term in the above multidimensional integration. Therefore, a wise selection of the position of the poles  according to the possible closure of the contour of integration in Fig.~\ref{fig4} must be done in order to avoid unnecessary analytic continuation. %Therefore,
%We are interested in the dependence on $(-\theta^2/k^2)$, which is equal to $(-k^2/\theta^2)$, because we want to have it in a region where the limits are possible without continuous analytical, 
We close the contour  to the left,    such that $|-\theta^2/k^2|>1$. Then, the poles are due to  
%\bea 
$\Gamma(\lambda_{12} + \lambda_{34} + \nu_{123} - \frac{D}{2} + \lambda_{5})$ alone,  and are located at
$\lambda_{5}=-\lambda_{12} -\lambda_{34}-\nu_{123} + \frac{D}{2} - n \;.$
Therefore
\bea 
I_{\lambda_{5}}&=&2\pi
i\sum_{n=0}^{\infty}\frac{(-1)^n}{n!}\bigg(-\frac{\theta^2}{k^2}\bigg)^{D/2-\lambda_{12} -\lambda_{34}-\nu_{123}-n} \frac{\Gamma(\lambda_{2} +
\nu_{3} +n)\Gamma\bigg(\lambda_{12}+\lambda_{34} + \nu_{123}
-\frac{D}{2} + n\bigg)}{\Gamma\big(\frac{D}{2} + \lambda_{12}
+n\big)\Gamma(2\lambda_{2} +\lambda_{1}+\nu_{13}+n)}\;. \nn
\eea
Hence, we can write
\bea
{\cal J}^{(0)}_{\nu_1,\nu_2,\nu_3}(\theta, m_{\pm})&=&\frac{(i)^D}{\prod_{i=1}^3\Gamma(\nu_{i})}\frac{(k^2)^{D/2
-\nu_{123}}}{(2 \pi
i)^3}\sum_{n=0}^{\infty}\frac{(-1)^n}{n!}\int_{-i\infty}^{+i\infty}\int_{-i\infty}^{+i\infty}\int_{-i\infty}^{+i\infty}d\lambda_{1}d\lambda_{2}d\lambda_{34}
\Gamma(-\lambda_{1}) \nn \\ && \times\Gamma(-\lambda_{2})\Gamma(-\lambda_{34})
\frac{\Gamma\big(\lambda_{2} +\nu_{3} +
n\big)\Gamma(\lambda_{12} +
\nu_{1})\Gamma\big(\lambda_{12} +\lambda_{34}+ \nu_{123} - \frac{D}{2} +
n\big)}{\Gamma\big(\frac{D}{2} + n
+ \lambda_{12}\big)\Gamma(n + \nu_{13} + \lambda_{1}+2\lambda_{2})}
\nn \\ && \times \Gamma(-\lambda_{2}-\lambda_{34} - \nu_{13} + \frac{D}{2})\bigg(\frac{p^2}{k^2}\bigg)^{\lambda_{1}}\bigg(\frac{q^2}{k^2}\bigg)^{\lambda_{2}}\bigg(-\frac{m^2}{k^2}\bigg)^{\lambda_{34}} \nn \\ && \times \bigg(-\frac{\theta^2}{k^2}\bigg)^{D/2 - \lambda_{12} -\lambda_{34}-\nu_{123} -n}\;. 
\eea 
Next, we integrate with respect to $\lambda_{34}$ closing the contour in Fig.~\ref{fig4} to the left, where $|-m^2/k^2|>1$. Naming $I_{\lambda_{34}}$ the relevant term, we have that 
\bea
I_{\lambda_{34}}&=&\int_{-i\infty}^{+i\infty}d\lambda_{34}\bigg(-\frac{m^2}{k^2}\bigg)^{\lambda_{34}}
\bigg(-\frac{\theta^2}{k^2}\bigg)^{D/2 -\lambda_{12}-\lambda_{34}-\nu_{123}-n}\Gamma(-\lambda_{34})
\Gamma\bigg(-\lambda_{2} - \lambda_{34} - \nu_{13} + \frac{D}{2}\bigg)
\nn\\
&&\times\Gamma\bigg(\lambda_{12} +
\lambda_{34} + \nu_{123}-\frac{D}{2} + n\bigg)\;. \nn
\eea 
The poles in $I_{\lambda_{34}}$ are due to $\Gamma(\lambda_{12} +\lambda_{34}+ \nu_{123}-\frac{D}{2} + n)$, 
and  are located at $\lambda_{34}=-\lambda_{12} -\nu_{123} + D/2 -n$. Thus 
\bea
I_{\lambda_{34}}&=&2\pi i\sum_{l=0}^{\infty}\frac{(-1)^l}{l!}\bigg(-\frac{\theta^2}{k^2}\bigg)^l\bigg(-\frac{m^2}{k^2}\bigg)^{-\lambda_{12} -\nu_{123} + D/2 -n-l}
\Gamma(\lambda_{1}+n+ l+\nu_{2})\nn \\ && \times\Gamma\bigg(\lambda_{12} +l+\nu_{123} - \frac{D}{2} + n\bigg)\;. \nn
\eea 
Then we can write
\bea
{\cal J}^{(0)}_{\nu_1,\nu_2,\nu_3}(\theta, m_{\pm})&=&\frac{(i)^D(k^2)^{D/2
-\nu_{123}}}{\prod_{i=1}^{3}\Gamma(\nu_{i})(2\pi
i)^2}\sum_{n,l=0}^{\infty}\frac{(-1)^n}{n!}\frac{(-1)^l}{l!}\bigg(-\frac{\theta^2}{k^2}\bigg)^l
\int_{-i\infty}^{+i\infty}\int_{-i\infty}^{+i\infty}d\lambda_{1}d\lambda_{2}\Gamma(-\lambda_{1})\nn
\\ && \times\Gamma(-\lambda_{2})\bigg(\frac{p^2}{k^2}\bigg)^{\lambda_{1}}\bigg(-\frac{m^2}{k^2}\bigg)^{\lambda_{12}-\nu_{123}+ D/2 -n-l} \bigg(\frac{q^2}{k^2}\bigg)^{\lambda_{2}}\Gamma(\lambda_{12}+\nu_{1})
\Gamma(\lambda_{2} +\nu_{3} + n)\nn \\ && \times \Gamma\bigg(\lambda_{12} +\nu_{123} -\frac{D}{2} +n+
l\bigg) \frac{\Gamma(\lambda_{1}+\nu_{2} + n +
l)}{\Gamma\big(\lambda_{12} + \frac{D}{2} + n
\big)\Gamma(2\lambda_{2} +\lambda_{1} + n + \nu_{13})}\;. 
\eea 
Now we integrate with respect to $\lambda_{2}$, but now closing the contour in Fig.~\ref{fig4} to the right, such that $|q^2/k^2|<1$. Calling $I_{\lambda_2}$ to the corresponding contribution, we have 
\bea
I_{\lambda_{2}}&=&\int_{-i\infty}^{+i\infty}d\lambda_{2}
\bigg(\frac{q^2}{k^2}\bigg)^{\lambda_{2}}\bigg(-\frac{m^2}{k^2}\bigg)^{-\lambda_{12}-\nu_{123}+D/2 -n-l}\Gamma(-\lambda_{2})\Gamma(\lambda_{12}+\nu_{1})\Gamma(\lambda_{2}+\nu_{3}+n)
\nn \\ && \frac{\Gamma\big(\lambda_{12}+n+l+\nu_{123}-\frac{D}{2}\big)}{\Gamma\big(\lambda_{12}+\frac{D}{2}+n\big)\Gamma(2\lambda_{2}+
\lambda_{1}+n+\nu_{13})}\;. \nn \eea 
There are poles in  this expression are due to $\Gamma(-\lambda_{2})$ and are located at
$\lambda_{2}=j$. Therefore
\bea
I_{\lambda_{2}}&=&
2\pi
i\sum_{j=0}^{\infty}\frac{(-1)^j}{j!}\bigg(\frac{q^2}{k^2}\bigg)^j\bigg(-\frac{m^2}{k^2}\bigg)^{-\lambda_{1}-j-\nu_{123}+D/2 -n-l}
\Gamma(\lambda_{1}+\nu_{1} + j)
\Gamma(n +j+\nu_{3})\nn \\ &&  
\times\frac{\Gamma\big(\lambda_{1}+n+l+j+\nu_{123}-\frac{D}{2}\big)}{\Gamma\big(\lambda_{1}+n+j+\frac{D}{2}\big)\Gamma(\lambda_{1}+n+2j+\nu_{13})}\;, \nn
\eea 
%\end{widetext}
%\newpage
and thus
%\begin{widetext}
\bea
{\cal J}^{(0)}_{\nu_1,\nu_2,\nu_3}(\theta, m_{\pm})&=&\frac{(i)^D(k^2)^{D/2-\nu_{123}}}{\prod_{i=1}^{3}\Gamma(\nu_{i})2\pi i}\sum_{n,l,j=0}^{\infty}
\frac{(-1)^n}{n!}\frac{(-1)^l}{l!}\frac{(-1)^j}{j!}\bigg(-\frac{\theta^2}{k^2}\bigg)^l\bigg(\frac{q^2}{k^2}\bigg)^j\Gamma(n+j+\nu_{3}) \nn \\ && \times \int_{-i\infty}^{+i\infty}d\lambda_{1}\Gamma(-\lambda_{1})
\bigg(\frac{p^2}{k^2}\bigg)^{\lambda_{1}}\bigg(-\frac{m^2}{k^2}\bigg)^{-\lambda_{1}-\nu_{123}+D/2-n-l-j}\Gamma(\lambda_{1} +n+l+\nu_{2})\nn \\ &&\times \Gamma(\lambda_{1} +j+\nu_{1})
\frac{\Gamma\big(\lambda_{1} +j+\nu_{123}-\frac{D}{2}+n+l\big)}{\Gamma\big(\lambda_{1}+n+j+\frac{D}{2}\big)\Gamma(\lambda_{1}+n+2j+\nu_{13})}\;. 
\eea 
Finally, we integrate with respect to $\lambda_{1}$ closing the contour to the right, such that $|p^2/k^2|<1$, and observing that the poles of the expression are due to 
$\Gamma(-\lambda_{1})$ and they are located at $\lambda_{1}=s$, 
we find
\bea
{\cal J}^{(0)}_{\nu_1,\nu_2,\nu_3}(\theta, m_{\pm})&=&\frac{(i)^D(-m^2)^{D/2-\nu_{123}}}{\prod_{i=1}^{3}\Gamma(\nu_{i})}\sum_{n,l,j,s=0}^{\infty}
\frac{(-1)^n}{n!}\frac{(-1)^l}{l!}\frac{(-1)^j}{j!}\frac{(-1)^s}{s!}\bigg(-\frac{k^2}{m^2}\bigg)^n\bigg(\frac{\theta^2}{m^2}\bigg)^l \nn \\ && \times\bigg(-\frac{q^2}{m^2}\bigg)^j
\bigg(-\frac{p^2}{m^2}\bigg)^s
\Gamma(\nu_{3}+n+j)
\Gamma(\nu_{2}+n+l+s)\Gamma(\nu_{1}+j+s)\nn \\ && \times
\frac{\Gamma\big(\nu_{123}-\frac{D}{2}+n+l+j+s\big)}{\Gamma\big(\frac{D}{2}+n+j+s\big)\Gamma(\nu_{13}+n+2j+s)}\;.
\eea 
Changing to Pochammer symbols according to 
\bea
\Gamma(\nu_{3}+n+j)&=&(\nu_{3};n+j)\Gamma(\nu_{3})\;, \nn \\
\Gamma(\nu_{2}+n+l+s)&=&(\nu_{2};n+l+s)\Gamma(\nu_{2})\;, \nn \\
\Gamma(\nu_{1}+j+s)&=&(\nu_{1};j+s)\Gamma(\nu_{1})\;, \nn \eea \bea
\Gamma\bigg(\nu_{123}-\frac{D}{2}+n+l+j+s\bigg)&=& \bigg(\nu_{123} -\frac{D}{2};n+l+j+s \bigg)\Gamma \bigg(\nu_{123}-\frac{D}{2}\bigg)\;, \nn \\
\Gamma(\nu_{13}+n+2j+s)&=&(\nu_{13};n+2j+s)\Gamma(\nu_{13})\;, \nn \\
\Gamma\bigg(\frac{D}{2}+n+j+s\bigg)&=&\bigg(\frac{D}{2};n+j+s\bigg)\Gamma\bigg(\frac{D}{2}\bigg)\;, \nn
\eea
we reach at the expression
\bea
{\cal J}^{(0)}_{\nu_1,\nu_2,\nu_3}(\theta, m_{\pm})&=&(i)^D(-m^2)^{D/2
-\nu_{123}}\frac{\Gamma\big(\nu_{123}-\frac{D}{2}\big)}{\Gamma\big(\frac{D}{2}\big)\Gamma(\nu_{123})}\sum_{n,l,j,s=0}^{\infty}\frac{(-1)^{n+l+j+s}}{n!l!j!s!}\bigg(-\frac{k^2}{m^2}\bigg)^n
\bigg(\frac{\theta^2}{m^2}\bigg)^l
\nn \\ && \times \bigg(-\frac{q^2}{m^2}\bigg)^j \bigg(-\frac{p^2}{m^2}\bigg)^s 
\frac{\big(\nu_{123}-\frac{D}{2};n+l+s+j\big)(\nu_{2};n+l+s)(\nu_{3};n+j)}{\big(\frac{D}{2};n+j+s\big)(\nu_{13};n+2j+s)}\nn \\ && \times (\nu_{1};j+s) \;, 
\eea 
or, equivalently 
\bea 
{\cal J}^{(0)}_{\nu_1,\nu_2,\nu_3}(\theta,m_{\pm})&=& \nn \\ && \hspace{-22 mm}(i)^D(-m^2)^{D/2-\nu_{123}}\frac{\Gamma\big(\nu_{123}-\frac{D}{2}\big)}{\Gamma\big(\frac{D}{2}\big)\Gamma(\nu_{123})}
\Upsilon_{3}\left[
\begin{array}{c}
\nu_{123}-\frac{D}{2},\nu_{2},\nu_{3},\nu_{1} \\ 
\frac{D}{2};\nu_{13}	
\end{array}
\Bigg|\frac{k^2}{m^2},\frac{\theta^2}{m^2},\frac{q^2}{m^2},\frac{p^2}{m^2}\right]\;, 
\eea
where $\Upsilon_{3}\left[
\begin{array}{c}
a_1,a_2,a_3,a_4 \\ 
b_1;b_2	
\end{array}
\Bigg|x_1,x_2,x_3,x_4\right]$  is the generalized Lauricella function of 4 variables. This is the general expression in arbitrary space-time dimensions. In our particular case, we take
$\nu_{1}=\nu_{2}=\nu_{3}=1$ and $D=3$, and Wick rotates to Euclidean space. We hence obtain
\bea
J^{(0)}(\theta, m_{\pm})&=&(m^2)^{-3/2
}\frac{\Gamma\big(3-\frac{3}{2}\big)}{\Gamma\big(\frac{3}{2}\big)\Gamma(3)}\sum_{n,l,j,s=0}^{\infty}\frac{(-1)^{n+l+j+s}}{n!l!j!s!}\bigg(\frac{k^2}{m^2}\bigg)^n
\bigg(\frac{\theta^2}{m^2}\bigg)^l\bigg(\frac{q^2}{m^2}\bigg)^j
\bigg(\frac{p^2}{m^2}\bigg)^s 
\nn \\ 
&&\times \frac{\big(3-\frac{3}{2};n+l+s+j\big)(1;n+l+s)(1;n+j)(1;j+s)}{\big(\frac{3}{2};n+j+s\big)(2;n+2j+s)}\;, \label{integral J}
\eea 
or
\bea 
J^{(0)}(\theta,m_{\pm})&=&(m^2)^{-3/2}
\Upsilon_{3}\left[
\begin{array}{c}
\frac{3}{2},1,1,1 \\ 
\frac{3}{2};2	
\end{array}
\Bigg|\frac{k^2}{m^2},\frac{\theta^2}{m^2},\frac{q^2}{m^2},\frac{p^2}{m^2}\right]\;. \nn \\
\eea
This completes the calculation of $J^{(0)}(\theta,m_{\pm})$. Next we check some particular limits. 

First we consider $\theta^2=m^2$. Then
\bea
J^{(0)}(m_{\pm})&=&(m^2)^{-3/2}\sum_{n,j,s=0}^{\infty}\frac{(-1)^{n+j+s}}{n!j!s!}\bigg(\frac{k^2}{m^2}\bigg)^n\bigg(\frac{q^2}{m^2}\bigg)^j\bigg(\frac{p^2}{m^2}\bigg)^s\frac{\Gamma(s + j+1)\Gamma(n+s+1)}{\Gamma(2n + 2j+ 2s + 3)} 
\times \nn \\ && \Gamma(n + j + 1) \Gamma\bigg(n + j + s + \frac{3}{2} \bigg)\;, 
\eea
that can be written as  
\bea
J^{(0)}(m_{\pm})&=&
\frac{\sqrt{\pi}}{4}(m^2)^{-3/2}\Phi_{3}\left[
\begin{array}{c}
\frac{3}{2},1,1,1 \\ 
3	
\end{array}
\Bigg|\frac{k^2}{m^2},\frac{q^2}{m^2},\frac{p^2}{m^2}\right]\;, \nn \\
\eea
where $\Phi_{3}\left[
\begin{array}{c}
a_1,a_2,a_3,a_4 \\ 
b_1	
\end{array}
\Bigg|x_1,x_2,x_3\right]$ is the generalized Lauricella function of three variables.
This expression coincides with the Eq. (37) obtained in Ref. \cite{davydychev} where, in our case, $\nu_{1}=\nu_{2}=\nu_{3}=1$ and $\nu_{1}\rightarrow \rho$, $\nu_{2}\rightarrow \mu$, $\nu_{3}\rightarrow \nu$. 

Another limit of interest is when $\theta=0$. This limit is achieved by returning to Eq.~(\ref{J0}), where we take directly $\theta=0$ and integrate using the Barnes theorem and integrating in accordance with above reasoning. Then,
\bea
J^{(0)}(m_{\pm})&=&(m^2)^{-3/2}\sum_{n,s,j=0}^{\infty}\frac{(-1)^{n+s+j}}{n!s!j!}\bigg(\frac{k^2}{m^2}\bigg)^n\bigg(\frac{p^2}{m^2}\bigg)^s\bigg(\frac{q^2}{m^2}\bigg)^j\Gamma(n + j+ 1) \nn \\ && \times \Gamma\bigg(n + s + j + \frac{3}{2}\bigg)\Gamma\bigg(j + \frac{1}{2}\bigg) \frac{\Gamma(s +j + 1)\Gamma(n+s + 1)}{\Gamma\big(n +s+j+ \frac{3}{2}\big)\Gamma(n + s+ 2j + 2)}
\eea
or
\bea
J^{(0)}(m_{\pm})&=&\sqrt{\pi}(m^2)^{-3/2}\Phi_{2}\left[
\begin{array}{c}
\frac{3}{2},1,1,1 ; \frac{1}{2} \\ 
\frac{3}{2}; 2	
\end{array}
\Bigg|\frac{k^2}{m^2},\frac{p^2}{m^2},\frac{q^2}{m^2}\right]\;, \nn \\
\eea
%\begin{widetext}
where $\Phi_{2}\left[
\begin{array}{c}
a_1,a_2,a_3,a_4 ; \tilde{a}_1 \\ 
b_1; \tilde{b}_1	
\end{array}
\Bigg|x_1,x_2,x_3\right]$ 
%\end{widetext}
is the three-parameter generalized Lauricella function. This expression is in agreement with
Eq.~(32) of the Ref.~\cite{davydychev} after the identification $\nu_{1}=\nu_{2}=\nu_{3}=1$ and $\nu_{1}\rightarrow \rho$, $\nu_{2}\rightarrow \mu$, $\nu_{3}\rightarrow \nu$. 

Finally, in the limit when $\theta^2\ll m^2$ in the Eq. (\ref{integral J}), we obtain
\bea
J^{(0)}(m_{\pm})&\approx&(m^2)^{-3/2}\sum_{n,j,s=0}^{\infty}\frac{(-1)^{n+j+s}}{n!j!s!}\bigg(\frac{k^2}{m^2}\bigg)^n\bigg(\frac{q^2}{m^2}\bigg)^j\bigg(\frac{p^2}{m^2}\bigg)^s \Gamma(j+s+ 1)\nn \\
&& \times \Gamma\bigg(n + j + s + \frac{3}{2}\bigg)\frac{\Gamma(n +j+ 1)\Gamma(n+s+1)}{\Gamma\big(n +j + s+ \frac{3}{2}\big)\Gamma(n +2j+s+ 2)} + \cdots \;.
\eea
\newpage
\begin{center}
{\bf 8. {\bf $J^{(1)}_{\mu}(\theta,m_\pm)$}}
\end{center}
We write $J^{(1)}_{\mu}(\theta,m_{\pm})$ in its most general form 
\begin{eqnarray}
{\it J}^{(1)}_{\mu}(\theta, m_{\pm})=\frac{\pi^2}{2}\left\{
k_{\mu}J_{A}(k,p)+p_{\mu}J_{B}(k,p)\right\} \;, 
\end{eqnarray}
that is symmetrical under the exchange of $k$ and $p$. Here
\begin{eqnarray}
J_{A}(k,p)&=&-\frac{1}{\Delta^2}   \Bigg\{  \bigg[ 
p^2 (k^2+ m_{\pm}^2 - \theta^2) -(k \cdot p)(p^2 + m_{\pm}^2 - \theta^2) \bigg] \frac{{\it J}_{0}(\theta,m_{\pm})}{2} \nn\\
  &&+( k \cdot p) \; \tilde{{\cal I}}(\theta^2,m_{\pm}^2,k^2)  - p^2 \; \tilde{{\cal I}}(\theta^2,m_{\pm}^2,p^2)  
  %\nn \\ && 
  +  \frac{1}{2} \, (p^2-k \cdot p) \; K_{0}(m_{\pm})
   \Bigg\} \; ,    \nonumber  \\
J_{B}(k,p)&=&J_{A}(p,k) \;,  
\end{eqnarray}
and $J_{0}(\theta,m_{\pm})=2/\pi^2J^{(0)}(\theta,m_{\pm})$, $K_{0}(m_{\pm})=2/\pi^2K^{(0)}(m_{\pm})$, $\tilde{{\cal I}}(\theta^2,m_{\pm}^2,p^2)={\cal I}(\theta^2,m_{\pm}^2,p^2)/\pi^2$. 
%We calculate ${\cal I}(\theta^2,m_{\pm}^2,p^2)$:
%\begin{eqnarray}
%{\cal I}(\theta^2, m_{\pm}^2, p^2)=\int d^3 w\frac{1}{[w^2 + \theta^2][(p - w)^2 +m_{\pm}^2]}\;. \nn \\
%\end{eqnarray}
%Using standard techniques, we find 
%\bea 
%{\cal I}(\theta^2, m^2_{\pm}, p^2)
%&=&2\pi^2 I(p,m_\pm+\theta)
%\eea 
The limits $\theta \rightarrow 0$, $m_{\pm} \rightarrow 0$ and $\theta \rightarrow m_{\pm}$ can be derived straightforwardly and show complete agreement with the findings of Refs.~\cite{raya, bashir}. \\

\begin{center}
{\bf 9. {\bf $J^{(2)}_{\mu\nu}(\theta,m_{\pm})$}}
\end{center}

In its most general form, we expand
\begin{eqnarray}
{\it J}^{(2)}_{\mu\nu}(\theta,m_{\pm})&=&\frac{\pi^2}{2}\Bigg\{
-\frac{\delta_{\mu\nu}}{3}K_0(m_{\pm})-\left(k_{\mu}k_{\nu}-\delta_{\mu\nu}\frac{k^2}{3}\right)
J_{C}(k,p) \nn \\ && - 
\left(p_{\mu}k_{\nu}+k_{\mu}p_{\nu}-\delta_{\mu\nu}\frac{2k\cdot p}{3}\right)J_{D}(k,p) \nn \\ &&
 - \left(p_{\mu}p_{\nu}-\delta_{\mu\nu}
\frac{p^2}{3}\right)J_{E}(k,p) \Bigg\} \;,
\end{eqnarray}

where
\begin{eqnarray}
J_{C}(k,p)&=&\frac{1}{\Delta^2}   
\Bigg\{ \bigg[ \frac{1}{2}(k\cdot p)(p^2 +m_{\pm}^2 -\theta^2) - p^2(k^2 +m_{\pm}^2-\theta^2) \bigg]\; 
J_A(k,p)  \nn \\ && -\frac{p^2}{2}(p^2+m^2_{\pm} -\theta^2) \; J_B(k,p) \;  + p_{\nu}\; \tilde{{\cal I}}_{\nu}(\theta^2,m^2_{\pm},k^2) \nn \\ && + (k \cdot p + p^2) \;\frac{K_{0}(m_{\pm})}{4}  \;
\Bigg\} \;,  
   \nonumber \eea
\bea
J_{D}(k,p)&=&\frac{3}{2\Delta^2}   
\Bigg\{ \hspace{2mm} \bigg[ (k\cdot p)(k^2 +m_{\pm}^2 -\theta^2) -\frac{1}{3}k^2(p^2 +m_{\pm}^2 -\theta^2)
\bigg] 
\; \frac{J_A}{2} \nn \\ && +  \bigg[(k\cdot p)(p^2 +m^2_{\pm}-\theta^2)  -\frac{p^2}{3}(k^2 +m_{\pm}^2 -\theta^2)
\bigg] \frac{J_B}{2}
 \nn \\ &&   - \frac{1}{4}\left[\frac{1}{3}(p-k)^2 + \frac{4}{3}(k\cdot p)\right]K_{0}(m_{\pm})\nn\\ &&+\frac{1}{3}[k_{\nu}\tilde{{\cal I}}_{\nu}(\theta^2,m^2_{\pm},k^2) +%\nn \\ && + 
p_{\nu}\tilde{{\cal I}}_{\nu}(\theta^2,m^2_{\pm},p^2)]
\Bigg\} \;,  
\nonumber \\
J_{E}(k,p)&=&J_{C}(p,k) \;,
\end{eqnarray}

and $\tilde{{\cal I}}_{\nu}(\theta^2,m^2_{\pm},p^2)={\cal I}_{\nu}(\theta^2,m^2_{\pm},p^2)/\pi^2$, being
\begin{eqnarray}
{\cal I}_{\nu}(\theta^2, m_{\pm}^2, p^2)=\int d^3 w\frac{w_{\nu}}{[w^2 + \theta^2][(p - w)^2 + m_{\pm}^2]}\;. \nn \\
\end{eqnarray}
This integral is readily found to be
\bea
{\cal I}_{\nu}(\theta^2, m^2_{\pm}, p^2)&=&\pi^2p_{\nu}\Bigg\{\frac{\theta-m_{\pm}}{p^2} + \frac{(m_{\pm}^2 - \theta^2 + p^2)}{p^2} I(p,m_\pm+\theta)\Bigg\}\;. \label{integral II}
\eea
The limits $\theta \rightarrow 0$, $m_{\pm} \rightarrow 0$ and $\theta \rightarrow m_{\pm}$ can be derived straightforwardly and show complete agreement with the findings of Refs.~\cite{raya, bashir}.\\

\begin{center}
{\bf 10. {\bf $L^{(1)}_{\mu}(\theta,m_{\pm})$}}
\end{center}
In order to integrate
\bea
L^{(1)}_{\mu}(\theta,m_{\pm})&=&\int d^3w\frac{w_{\mu}}{w^2[w^2 +\theta^2][(p - w)^2 +m_{\pm}^2][(k - w)^2 +m_{\pm}^2]}, 
\eea
we use the identity
\be
\frac{1}{w^2[w^2 + \theta^2]}=\frac{1}{\theta^2}\left[\frac{1}{w^2} -\frac{1}{w^2 + \theta^2}\right]\;. 
\ee
Thus, we can write
\be
L^{(1)}_{\mu}(\theta,m_{\pm})=\frac{1}{\theta^2}[J^{(1)}_{\mu}(m_{\pm})-J^{(1)}_{\mu}(\theta,m_{\pm})]\;.
\ee 

\begin{center}
{\bf 11. {\bf $L^{(2)}_{ \mu \nu}(\theta,m_{\pm})$}}
\end{center}
Similarly, we can express
\be
L^{(2)}_{\mu\nu}(\theta,m_{\pm})=\frac{1}{\theta^2}[J^{(2)}_{\mu\nu}(m_{\pm})-J^{(2)}_{\mu\nu}(\theta,m_{\pm})]\;.
\ee

\vfil\eject

\end{document}